# An Interpretable Closed-Loop Intelligent Tutoring System for Multimodal Affective Feedback in Asynchronous Presentation Training

Hung-Yue Suen and Kuo-En Hung

***Abstract*— This paper presents an interpretable closed-loop Intelligent Tutoring System (ITS) that supports feedback-guided practice for developing on-camera oral presentation skills at scale. The system operationalizes a seven-dimensional Behaviorally Anchored Rating Scale (BARS) and implements a three-layer interpretable feedback architecture that connects rubric-aligned multimodal scoring, audience-perceived expressive diagnostics, and retrieval-augmented conversational coaching to support deliberate practice. Built on an XGBoost backbone, the ITS maps multimodal inputs (facial, vocal, textual, and oculomotor features) into evidence-based feedback that can be traced back to observable performance cues. Trained on 10,360 Massive Open Online Course (MOOC) video segments, the system achieved rubric-aligned scoring with performance levels comparable to expert ratings ($R^2$ = 0.48–0.61, ρ = 0.69–0.78, MAE = 0.43–0.57). In a pre–post validation study with 204 adult learners over a 30-day practice window, participants demonstrated significant improvements across all seven BARS dimensions (Cohen's d = 0.39–0.90), with practice frequency showing a strong positive association with posttest performance after controlling for baseline scores and demographics. The results demonstrate how multimodal analytic outputs can be systematically transformed into observable behavioral change through an integrated feedback architecture, advancing explainable and pedagogically grounded ITS design for performance-based competencies.**

## I. INTRODUCTION

Presentation on camera has emerged as a critical competency in modern educational and professional landscapes. It acts as the cornerstone for a wide range of digital activities, from delivering MOOCs and video lectures to navigating virtual job interviews and disseminating knowledge online. As suggested by recent literature, this skill has transcended niche, task-specific contexts to become a fundamental requirement for broader success [1]

This work was supported by the National Science and Technology Council (NSTC), Taiwan, under Grants NSTC 114-2410-H-003-019-MY2 and NSTC 114-2622-H-003. All authors contributed equally to this work.
(*Corresponding author: Hung-Yue Suen.*)
Hung-Yue Suen is with Technology Application and Human Resource Development, National Taiwan Normal University, Taipei, Taiwan (ORCID: 0000-0002-6796-2031; e-mail: collin.suen@ntnu.edu.tw).
Kuo-En Hung is with Technology Application and Human Resource Development, National Taiwan Normal University, Taipei, Taiwan (ORCID: 0000-0003-2091-2747; e-mail: kuanntw@gmail.com).

However, opportunities to develop this skill remain uneven. Many users still rely on self-recording and informal self-review, which may increase awareness but rarely provide concrete guidance for improvement, and more critically, fail to ensure that feedback is translated into observable behavioral change. [2], [3]. Approaches that involve peer feedback or expert coaching can offer more detailed and context-sensitive input, yet these methods are time-consuming, costly, and difficult to scale consistently [4]. For most students and working professionals, access to systematic and objective support for improving on-camera delivery is still limited [5].

At the same time, research in machine learning (ML) and multimodal learning analytics has shown that observable cues from facial movement, vocal prosody, and spoken content are informative for understanding how presentations are perceived [2], [6]. Building on this line of work, several automated feedback systems have explored emotion-related signals and real-time coaching as ways to support speaker development [2], [7]. More recently, large language models (LLMs) combined with retrieval-augmented generation (RAG) have made it possible to generate individualized textual feedback that addresses aspects such as structure, word choice, and rhetorical flow [4].

Despite these advances, existing systems tend to address only parts of the problem, particularly by focusing on assessment and feedback generation without ensuring that such feedback leads to measurable behavioral change. Analyses of delivery behaviors, including prosody and facial expression [2], [6], [9], are often developed separately from language-based content coaching [4]. Therefore, users typically receive feedback on either how they speak or what they say, but rarely within a single learning process [1], [10]. In addition, many systems emphasize predictive performance while offering limited insight into why a score was produced or how a learner might improve [2], [3]. Finally, empirical evidence on learning effects remains relatively sparse. Although commercial tools are increasingly available [11], studies that combine rubric-aligned assessment with measurable improvement over time are still uncommon [3], [5], [6].

To address these limitations, this study reframes the intelligent tutoring system (ITS) as a closed-loop learning process in which learners repeatedly practice and adjust their behavior. Instead of positioning assessment as an endpoint, the system uses multimodal analytic results to generate feedback that learners can act on in subsequent attempts. This approach is consistent with recent work that extends ITS beyond cognitive skill training to observable expressive aspects of performance, as discussed in research on affective tutoring systems [12]. While existing multimodal and real-time learning systems have advanced in behavioral prediction, feedback delivery, and immediate learner guidance, the explicit mapping from multimodal analytic evidence to rubric-aligned learner interpretation and observable behavioral change remains underspecified in the literature. This study directly addresses this gap by operationalizing multimodal analytics into a feedback-driven learning mechanism that links rubric-aligned interpretation to actionable behavioral adjustment.

Within the proposed system, facial, vocal, textual, and eye-movement features are mapped onto Behaviorally Anchored Rating Scale (BARS) dimensions using an interpretable machine learning model based on XGBoost [13]. The resulting scores are presented through a visual dashboard and complemented by a conversational tutoring interface that translates analytic results into practice-oriented feedback. This combination allows learners to review performance, reflect on specific aspects of delivery, and decide how to adjust their next attempt.

This work makes the following major contributions:

1) Rubric-Aligned Multimodal Assessment: We operationalize educational rubrics as multimodal behavioral indicators (facial expression, vocal prosody, gaze, linguistic content), enabling an interpretable automated assessment that aligns with expert evaluation criteria.
2) Closed-Loop, Interpretable ITS Architecture: Rather than introducing a new predictive model, we propose a verifiable ITS framework that links multimodal assessment, diagnostic interpretation, and conversational feedback within a closed-loop structure. This design enables the system to support and evaluate learning processes through iterative cycles of feedback and behavioral adjustment.
3) Field-Based Evidence of Practice-Linked Learning: Using 10,360 MOOC video segments and a 30-day field study (N = 204), we show that performance gains are systematically associated with documented practice frequency and accompanied by changes in expressive behavior across dimensions.

This work develops a closed-loop feedback infrastructure for asynchronous, on-camera presentation refinement. It introduces a tightly coupled multimodal–LLM architecture that restructures analytic representations to enable direct and verifiable mapping from multimodal inference to actionable feedback for behavioral adjustment. By integrating multimodal assessment, interpretable analytics, and a conversational feedback interface under this design, the system supports structured cycles of performance review and iterative behavioral adjustment. This establishes a learning mechanism in which analytic evidence is systematically translated into learner interpretation and subsequent behavioral change.

## II. Related Work and Background

### A. ITSs for Presentation Skills

ITSs have focused more on cognitive domains such as math, science, and programming [14]. Meta-analyses have demonstrated that ITSs exceed traditional tutoring and static e-learning by moderate to large margins [15], [16]. Modern multimodal analysis and NLP have opened the possibility that ITSs can address skills that involve oral presentation, whereby presentation is dependent upon delivery and content equally [17]. Effective presentation requires coordinated verbal and nonverbal expressions [18].

Several studies have applied ITSs to presentation training. Classroom studies show multimodal feedback improves delivery [1], and prior multimodal analytics frameworks for automated feedback provide open-source pipelines for reproducible analysis [2]. Predictive models using speaker and audience sensing achieve reliable automated assessment [5], and transfer learning extends these models across contexts [6]. Recent work explores emotion recognition [7] and AI-clone-based self-reflection [8] as complementary training mechanisms. At the design level, principles for oral presentation skills have been synthesized in educational research [10], and extensive studies validate that video style consistently affects learner engagement in online environments [9]. Recent research delves into the potential of generative AI and RAG to convert analytic traces into contextualized coaching, highlighting the intersection of tutoring and coaching paradigms.

Self-regulation theory suggests that performance feedback alone can prompt learners to adjust their behaviors [19]. Nonetheless, many studies use perception-based metrics rather than rubric-aligned assessments, limiting interpretability and pedagogical validity [14]. Empirical validation is also constrained, as many studies are either short-term or corpus-based, with a scarcity of pre–post designs to assess enduring learning effects [3], [12]. Evidence from randomized trials suggests that supplementary practice sessions can further augment the development of presentation skills [20]; however, such methodologies are still infrequent in online self-directed learning contexts [9]. Addressing these gaps requires systems that integrate multimodal assessment with validated rubrics (e.g., BARS [18]) and demonstrate learning outcomes through rigorous empirical evaluation.

### B. Multimodal Analytics for Presentation Skills

Oral presentation is fundamentally a multimodal task, as the effectiveness of presentation is contingent upon vocal prosody, facial expressions, eye movements, and bodily actions across time dimensions [18]. Methodologies are being used to automatically mine and fuse multiple modal cues [2], [21]. Studies involving classroom settings provide evidence of enhanced presentation effectiveness when multimodal feedback is used [1], while predictive models involving speaker-and-audience sensing further enhance presentation skills [5]. Transfer learning is used for making presentation systems more robust in different settings [6], while methodologies in affective computing have broadened multiple modal recognition in educational applications [7]. Tutorial video applications involving AI-clone systems illustrate self-reflection analysis in training procedures [8].

Concerning the MOOC measures, the restrictions imposed by the conditions of the video recordings, for instance, single-camera recordings, affect the extraction of features and complex analysis of automated delivery [9]. Computer vision techniques now make it possible to calculate eye gaze based on webcam recordings without any trackers. Appearance-based techniques exploit eye gaze and head pose [22]. Methods based on facial landmarks (e.g., 33, 133, 362, 263) detect approximate representations of eye gaze [23]. Progress made in acoustic and lexical-semantic embeddings improves multimodal representation, exemplifying the complementarity between speech characteristics and framing concerning eye expressiveness [5, 6, 17, 24].

Ensemble models such as XGBoost provide a stable way of multimodal feature fusion with easily interpretable feature importance values [24]. Gradient boosting decision trees allow faster prediction and more interpretability compared to the use of deep neural networks, both of which are necessary for educational implementations [25], [26]. XGBoost frameworks can integrate acoustic, text, and image features effectively for large-scale analysis of MOOCs by achieving accuracy, interpretability, and validity simultaneously through XGBoost [25].

In this study, interpretability is treated as a pedagogical requirement rather than solely a modeling property. The use of XGBoost enables direct mapping between multimodal features and rubric-aligned feedback, allowing learners to identify specific aspects of their performance that require improvement, which is less directly supported by deep learning architectures. Accordingly, in this implementation, multimodal features derived from acoustic, textual, and visual signals are extracted independently and combined at the feature level prior to model training. This feature-level integration corresponds to an early fusion strategy. We adopted this approach to allow the model to capture interactions across modalities, such as how vocal delivery, facial expression, and linguistic content jointly contribute to presentation performance. In contrast, late fusion methods combine modality-specific predictions at the decision level, which may overlook such cross-modal relationships. Given the goal of linking multimodal signals to interpretable, rubric-aligned feedback, early fusion provides a more suitable design for capturing integrated behavioral patterns. This is particularly important in presentation assessment, where performance emerges from the integration of multiple modalities rather than from isolated unimodal signals.

### *C. Automated Feedback for Presentation Skills*

ITSs, in essence, go beyond automated scores since they combine assessment, feedback, and adaptation [14]. Users in presentation training benefit from both assessment and feedback, which helps them improve [2]. The effectiveness of feedback relies on the users' perception regarding the value and accuracy of feedback [27]. Current systems combine dashboards and dialogue systems for converting analytics data into contextualized feedback [28]. Reviews emphasize that investigations into soft-skill ITSs should prioritize quantifiable learning outcomes over subjective assessments [20].

Affective computing techniques have now integrated facial, speech, and linguistic features with performance evaluation [7], [30]. Automated feedback systems integrate speech as well as non-speech cues for automated evaluation and feedback, achieving a performance equivalent to expert judgments [31].

Self-regulation theories are linked to feedback that enables self-regulation and reflection of performance by learners [19]. The AutoTutor project illustrates that dialogue systems contribute to deeper-level understanding by using dialogue to scaffold reflection processes [32]. Based on such theories, we developed a conversational AI tutoring system by using multimodal feedback for reflection to extend sociocognitive tutoring systems for performance skills. However, there is a scarcity of empirical evidence for affective feedback in soft skills learning [33].

## III. METHODOLOGY

### *A. Research Framework*

**Fig 1** presents the overall research framework, which follows a dual-track design that integrates large-scale model development with learner-centered validation. On the modeling track, presentation performance is operationalized using a behaviorally anchored rating scale (BARS) that decomposes oral presentation quality into observable behavioral dimensions. The deployed system is examined in an authentic learning context to assess whether system use is associated with performance improvement over time.

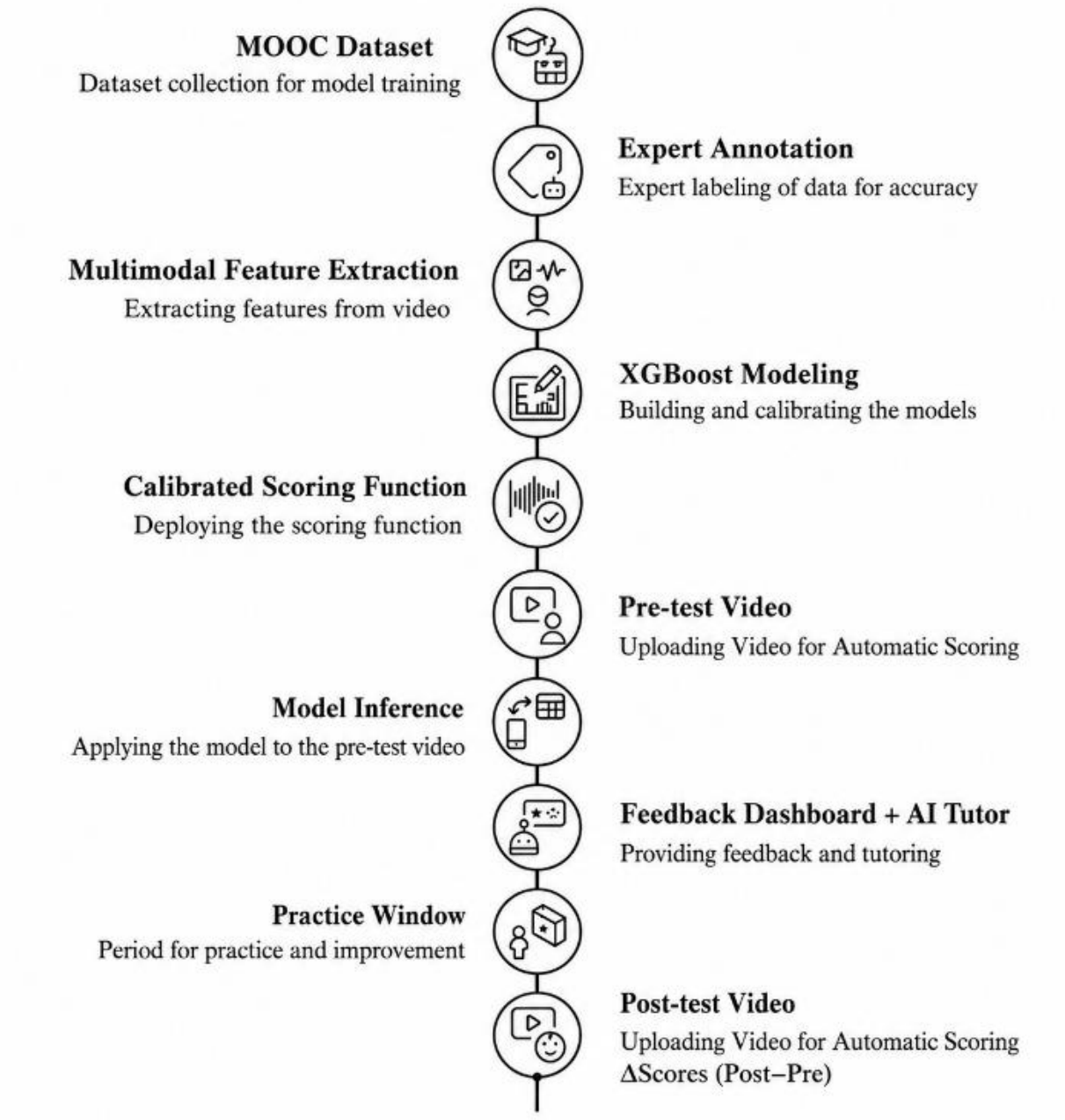

**Fig. 1**. Research framework

The framework adopts a seven-dimensional BARS structure to support both analytic assessment and holistic judgment. Six analytic dimensions (Topic, Content, Clarity, Voice & Talk,

Eye Contact, and Nonverbal Expressiveness) capture distinct, observable aspects of delivery, enabling fine-grained behavioral diagnosis. In contrast, the Overall Rating represents a global audience impression that reflects how multiple cues are integrated during impression formation. Prior research suggests that such holistic judgments may include emergent qualities that are not reducible to individual components alone [34], making the Overall Rating conceptually related to, yet distinct from, the analytic dimensions.

Together, these dimensions form the basis of an interpretable scoring architecture that supports scalable feedback generation. The system closes the feedback loop by translating rubric-aligned analytic outputs into actionable guidance through a multimodal dashboard and a conversational tutor. Learner validation is conducted through a pre–post study design in which presentation performance before and after system use is compared to examine instructional effectiveness.

Overall, the framework integrates rubric-based multimodal assessment, interpretable machine learning, and learner-centered evaluation into a coherent intelligent tutoring system for developing on-camera oral presentation skills.

### *B. Dataset and Annotation*

The modeling data consisted of presentation videos gathered from 120 independent instructors by collaborating with three Taiwanese MOOC platforms, namely Hahow, TibaMe, and TBK. Anonymization of all data was done by the collaborating platforms before being shared, in accordance with the data protection policies of their respective institutions. The data included instructors of both genders, with 72% male and 28% female, across various subject areas, of which 82% were from the social sciences and 18% from the sciences.

In each video, the first 2 seconds were removed because the speakers had not yet stabilized their speech at the beginning of the recording session [34]. The remaining portion of each video was then analyzed using voice activity detection (VAD) to determine speech-active portions, and the remaining portions not involving speech, such as transitions and long pauses, were eliminated from the analysis. The remaining speech-active portions of the videos were then segmented into 10,360 clips of 2 seconds each, using a sliding window with a 1-second stride, averaging 86 segments per video.

About 30 experts in online presentation pedagogy were recruited for annotation. Each instructor video was rated holistically by three independent annotators at the video level, and the resulting scores were used to supervise model training on the corresponding segments.

An adapted seven-item version of BARS was employed for annotation purposes (**Table I**). The BARS rubric was developed from the original scale described in [5] after a series of workshops with experts, aimed at making each dimension concrete enough to be observed in a single-camera instructor video.

Each dimension was anchored at levels 1, 3, and 5, with intermediate ratings permitted. Inter-rater reliability was assessed using intraclass correlation coefficients based on a one-way random-effects model for average ratings (ICC(1,k)). Reliability estimates ranged from 0.78 to 0.89 across dimensions, indicating strong consistency among expert raters.

**Table I**. BARS for Presentation Skills

| Dimension | Description | Anchors |
|---|---|---|
| **Topic** | How interestingly did the speaker present the topic? | **1:** The Topic was irrelevant or unengaging, failed to capture attention.<br>**3:** The Topic was moderately interesting, held partial audience attention.<br>**5:** The Topic was highly relevant and consistently engaging. |
| **Content** | How technically sound and substantive was the presentation content? | **1:** Content was inaccurate, superficial, or lacked depth.<br>**3:** Content was moderately accurate and somewhat developed.<br>**5:** Content was accurate, comprehensive, and well supported. |
| **Clarity** | How clearly did the speaker convey the essential ideas? | **1:** Ideas were unclear and disorganized, difficult to follow.<br>**3:** Ideas were partly clear but inconsistently expressed.<br>**5:** Ideas were consistently clear, logical, and easy to follow. |
| **Voice & talk** | How engaging and expressive was the speaker's voice and speech delivery? | **1:** Monotone or unclear voice; failed to engage.<br>**3:** Moderately engaging with some variation; uneven clarity.<br>**5:** Highly engaging, clear, and expressive throughout. |
| **Eye contact** | How effectively did the speaker establish eye contact with the audience? | **1:** Rarely or never established eye contact.<br>**3:** Occasional but not sustained eye contact.<br>**5:** Consistent and confident eye contact that engaged the audience. |
| **Nonverbal expressiveness** | How appropriate and supportive were the speaker's facial expressions and head movement? | **1:** Facial expressions or head movements were absent, distracting, or inappropriate.<br>**3:** Some facial expressions or head movements were appropriate but inconsistent.<br>**5:** Facial expressions and head movements consistently supported the expression. |
| **Overall Rating** | How would you evaluate the speaker's overall presentation performance? | **1:** Inferior overall presentation; major weaknesses across dimensions.<br>**3:** Adequate overall presentation; some strengths but limited effectiveness.<br>**5:** Excellent overall presentation; coherent, polished, and impactful |

### *C. Multimodal Analytics*

This study identified four categories of speaker features—facial dynamics, oculomotor signals, acoustic prosody, and semantic content—from asynchronous instructional videos. Each modality was processed and aligned independently for each segment to create a multimodal representation for predicting on-camera oral presentation scores [35].

1) **Facial features**

We used MediaPipe FaceMesh to look at how expressive people's faces were. It finds 478 three-dimensional facial landmarks per frame. Seven anatomical regions were used to group landmarks (for example, eyes, lips, and eyebrows). Then, geometric and dynamic descriptors were calculated, such as the mean and standard deviation of landmark positions, inter-point distances, angular relations, and inter-frame displacement velocity. A 2-second window (60 frames at 30 fps) kept short-term temporal dynamics, which led to a 3,780-dimensional facial feature vector for each segment [36].

2) **Oculomotor features**

We used the same set of landmarks to estimate eye-tracking and gaze behavior, focusing on points in the eye region. Some of the indicators were average fixation duration, saccadic transitions, and dispersion statistics. These were compiled into a seven-dimensional vector, enhancing facial dynamics with attentional signals [37].

3) **Acoustic features**

We processed audio signals ahead of time to get strong indicators of how well the voice was delivered. Acoustic features were extracted at two levels. First, we identified a representative 30-second segment per video exhibiting stable prosody (minimum variance in F0 and energy), serving as a speaker-level acoustic signature. Second, we extracted frame-level features (MFCCs, pitch, formants, spectral descriptors [38]) from each 2-second segment. The speaker signature was concatenated with segment-level features to provide both global speaker characteristics and local prosodic variation.

4) **Textual features**

We used Whisper-large-v2 ASR to write down what was said. The multilingual MiniLM model from SentenceTransformer turned transcripts into sentence embeddings that captured both semantic and affective context. We averaged the segment-level embeddings to get fixed-length textual representations [39].

5) **Modeling**

Low-level continuous features were extracted from four modalities: facial, oculomotor, acoustic, and textual cues. Seven XGBoost regression models were trained, including one model for the Overall Rating and six models corresponding to the remaining BARS dimensions. For each presentation, feature vectors from the four modalities were concatenated and used as input to the regression model associated with the target dimension. The Overall Rating model served as the primary outcome, while the other models generated dimension-level diagnostics for the system.

Model hyperparameters were tuned using Bayesian optimization within a predefined search space [40]. Model robustness was evaluated through speaker-independent cross-validation, such that no speaker appeared in both the training and testing sets, thereby avoiding identity leakage [41]. Model performance was assessed using the coefficient of determination ($R^2$) and mean absolute error (MAE). Spearman's rank correlation coefficient ($\rho$) was additionally reported to examine rank-order agreement between model predictions and expert ratings [42].

6) **Expressive Emotions Estimation**

We made separate emotion recognition modules to make affective descriptors that go along with the XGBoost predictions. Facial frames were taken every 0.5 seconds and sorted by a Convolutional Neural Network (CNN)-based recognizer [36]. The frame-level posteriors were then averaged into facial emotion probability vectors. Following [43] and [44], audio segments were run through a Speech Emotion Recognition (SER) pipeline to get vocal emotion probabilities. Following [44], speech transcripts were run through a valence-arousal lexicon to get textual affect distributions. We normalized all modality-specific probabilities per video and kept them as descriptive information for feedback visualization. These affective representations are employed in the feedback interface to assist users in deciphering the expressive and emotional inclinations that inform their BARS scores [2], [12].

### *D. Automated Feedback Ecosystem*

For the feedback provided to be more related to learning than a simple scoring system, the ITS is structured as a closed-loop feedback system with three distinct layers. These layers are set up to mirror how users perceive the outcome of the performance, the underlying causes of the performance, and the means of improvement.

Beginning with the first layer, the value of the system lies in highlighting the performance of the presentation related to the primary focus of instruction, and it emphasizes the Overall Rating of the performance, in addition to the six other BARS measures. This layer supports performance outcomes and helps the user pinpoint the information in the presentation that needs to be addressed.

The second layer provides emotion-aware expressive diagnosis by using observable, audience-facing expressive behaviors, including facial, vocal, verbal, and lexical emotive expressions, as behavioral signals rather than direct evidence of the speaker's internal emotional states. Consistent with observer-based affect perception frameworks [34], these cues are treated as behavioral signals that provide interpretable information about presentation delivery and are linked to the BARS-based evaluation structure. Specifically, facial expressions support interpretation of Nonverbal Expressiveness, vocal and verbal expressions support interpretation of Voice & Talk, and lexical expressions help contextualize dimensions such as Topic, Content, and Clarity. The system uses these indicators as explanatory cues that help learners understand why specific presentation dimensions may be underperforming.

At the third layer, a Conversational AI Tutor supports the dashboard through an interactive process to resolve queries and doubts created through the diagnostic displays in the dashboard results. By using the learner's results in the dashboard and segment levels, the tutor enhances both analytical guidance and improvement steps.

1) **Multimodal Feedback Dashboard**

The multimodal feedback dashboard offers an integrated view that combines presentation performance with expressive features available in multimodal feedback. The interaction process starts with a summary report (**Fig. 2**), which shows individualized textual feedback that combines presentation performance with important expressive features related to instructional improvement.

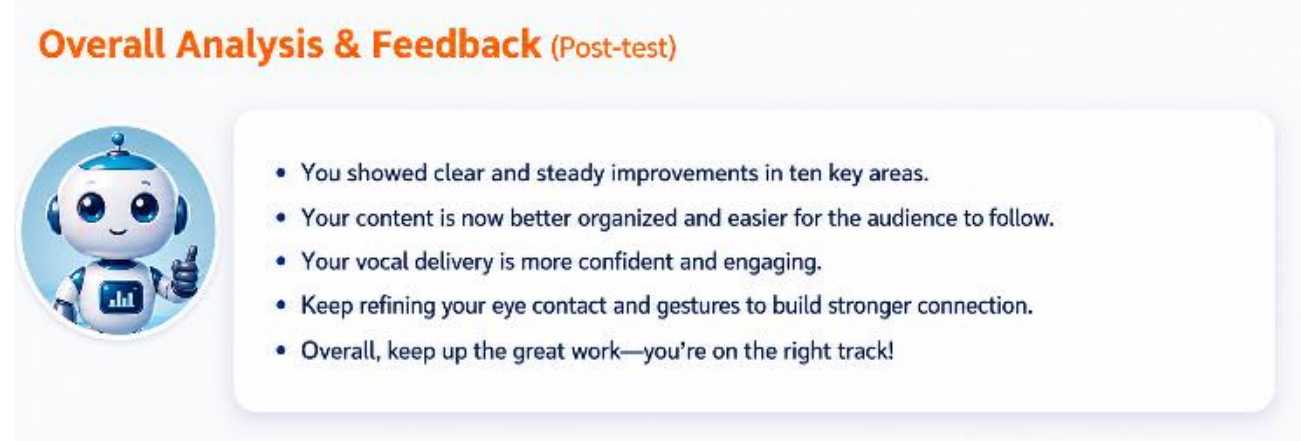

**Fig. 2. Summary Report for Oral Expressions**

This summary offers a translation of analytical outcomes into brief and general comments about the presentation skill level of the user as well as tendencies in terms of presentation style, pinpointing areas that require further thought and practice. This summary not only facilitates a quick understanding on the part of the learners about the impression of the presentation but also facilitates a subsequent analysis in other modules.

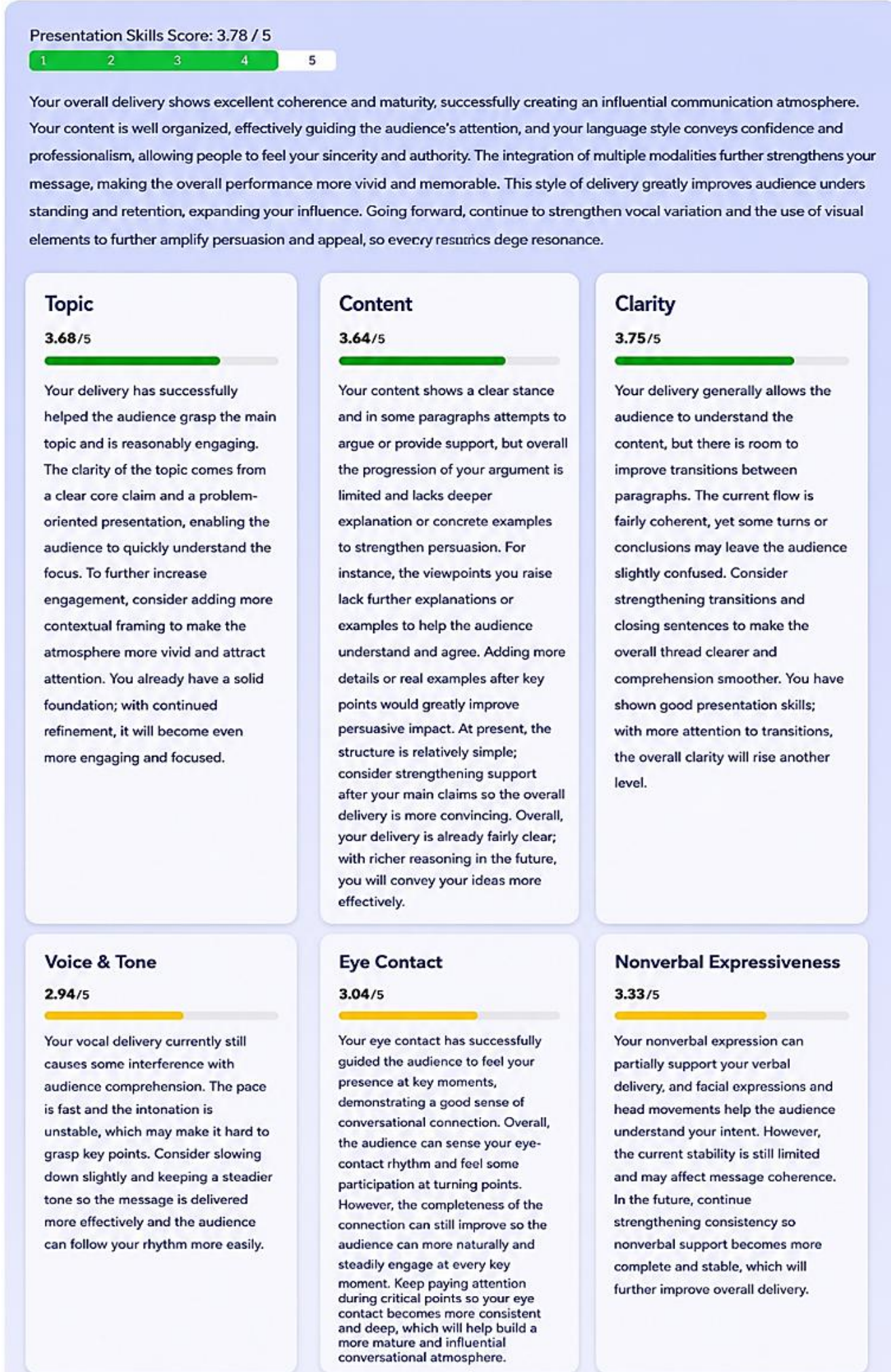

**Fig. 3. Presentation Scores**

After the summary, a module highlights the performance of the presentation through presentation score cards that focus on the Overall Rating as well as six BARS-related sub-dimensions (**Fig. 3**). BARS sub-dimensions are comprised of essential parts of oral presentation skills that serve as a framework for assessing on-camera presentation skills.

Through such visualizations, the dashboard allows learners to see how their current presentation measures up against expected standards set by the rubric, as well as pinpoint which factors of performance have the greatest negative impact on general expressiveness. As opposed to being focused on raw scores, this module was created with the intention of assisting self-regulated learning by helping learners pinpoint where concentrated study will provide the best results.

The dashboard introduces an emotion-aware expressive layer aligned with the BARS-based evaluation structure. In this system, affective signals are not designed to capture the speaker's internal emotional states. Rather, they capture expressive cues that are perceivable by an audience and that shape how the presentation is received, consistent with observer-based affect perception research in which brief behavioral samples reliably predict interpersonal impressions [34]. These cues are operationalized as observable expressive behaviors (e.g., vocal modulation, gaze, and facial movement) that provide interpretable indicators of how multimodal features relate to presentation performance.

The facial and vocal modules examine how learners express themselves through six basic emotions—anger, disgust, fear, happiness, sadness, and surprise. By presenting emotion distributions over time, the dashboard allows learners to observe whether their emotional expression remains stable, becomes over-intensified, muted, or fluctuates inconsistently during delivery. Learners can click on individual slices to access representative video or audio segments, enabling direct review of the facial or vocal behaviors that most strongly contribute to the observed affective patterns.

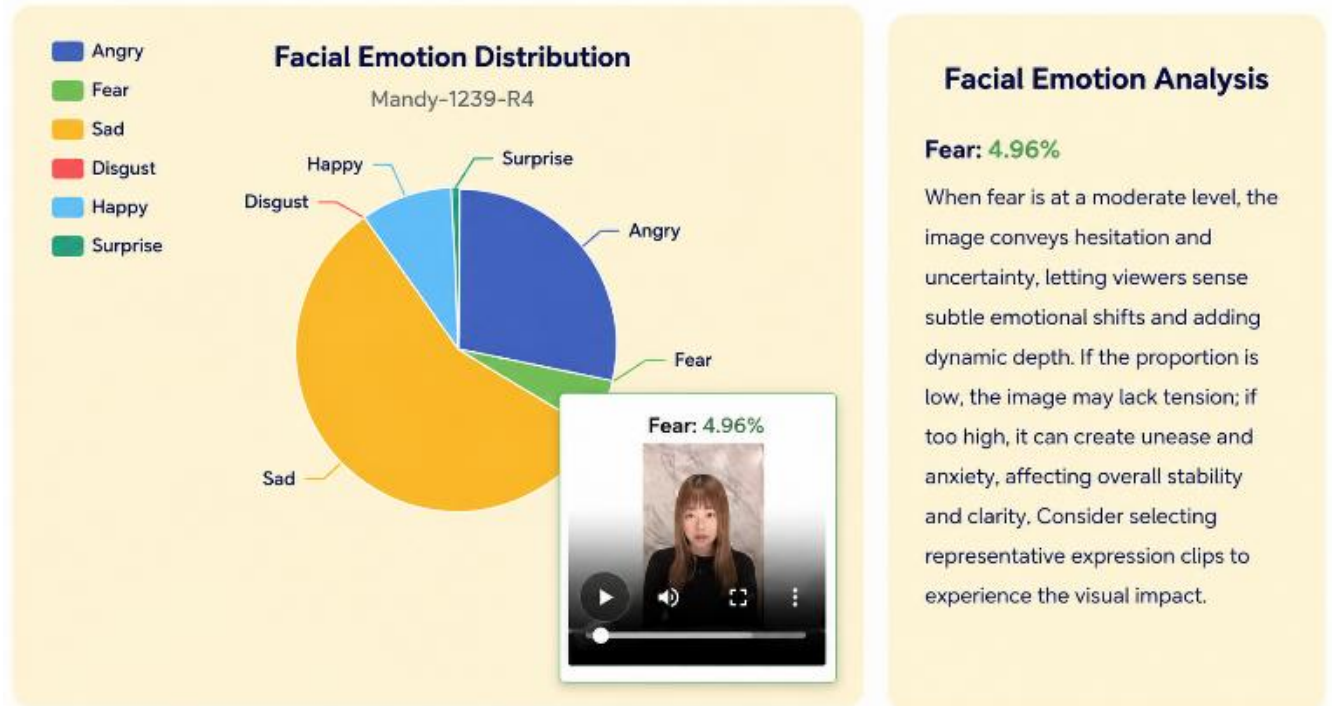

**Fig. 4. Facial Emotive Expressions**

Within the facial module (**Fig. 4**), the dashboard presents segment-level facial emotive expressions derived from video footage, summarizing the relative proportions of different emotional states observed across the presentation. By visualizing how frequently specific emotions occur and how they are distributed across segments, the module enables learners to consider whether their facial expression patterns are congruent with communicative intent and how these affective tendencies may influence Nonverbal Expressiveness and Eye Contact. By linking emotion proportions to observable facial behavior, this module supports reflection on audience-facing presence and helps learners evaluate whether their facial delivery conveys appropriate engagement and expressiveness.

Within the vocal module (**Fig. 5**), the dashboard presents a segment-level analysis of speech audio that summarizes the relative proportions of six basic emotions expressed through vocal delivery across the presentation. By visualizing how frequently each emotional state is conveyed in speech, the module allows learners to examine whether vocal expression is characterized by appropriate emotional variation, excessive tension, or emotional flatness. Given the critical role of vocal dynamics in perceived speech quality, this representation supports reflection on how emotion-related vocal patterns may influence Voice & Talk, particularly in terms of expressiveness, engagement, and perceived confidence. Linking emotion proportions to specific speech segments further enables learners to revisit representative excerpts and refine techniques of affective vocal modulation.

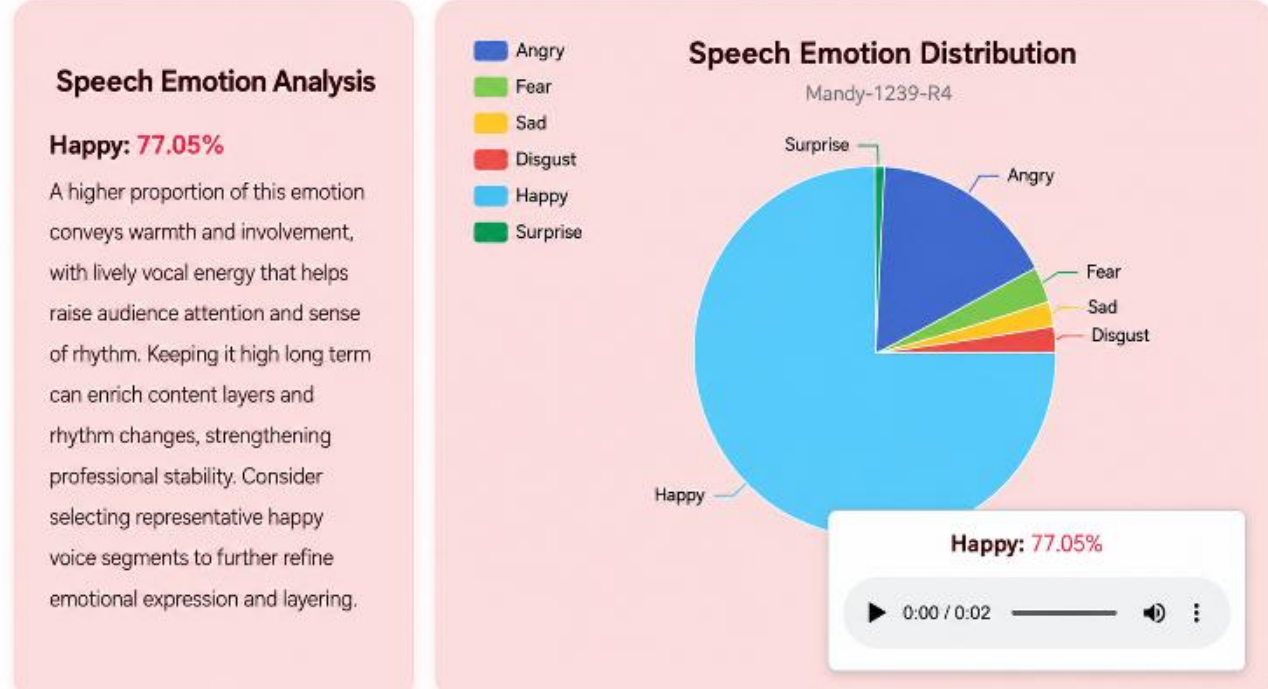

**Fig. 5. Vocal Emotive Expressions**

The textual module extends emotion-aware diagnostics to verbal expression by focusing on affective meaning conveyed through linguistic structure, tone, and rhetorical clarity. As shown in **Fig. 6**, transcribed speech is analyzed using a valence–arousal framework with four affective orientations: positive–high, positive–low, negative–high, and negative–low [44]. By summarizing the relative distribution of these affective orientations across the presentation, the visualization enables learners to understand how lexical choices shape perceived vocal expressiveness and supports more deliberate adjustment of wording and provides a complementary interpretive channel for refining Voice & Talk at the linguistic level.

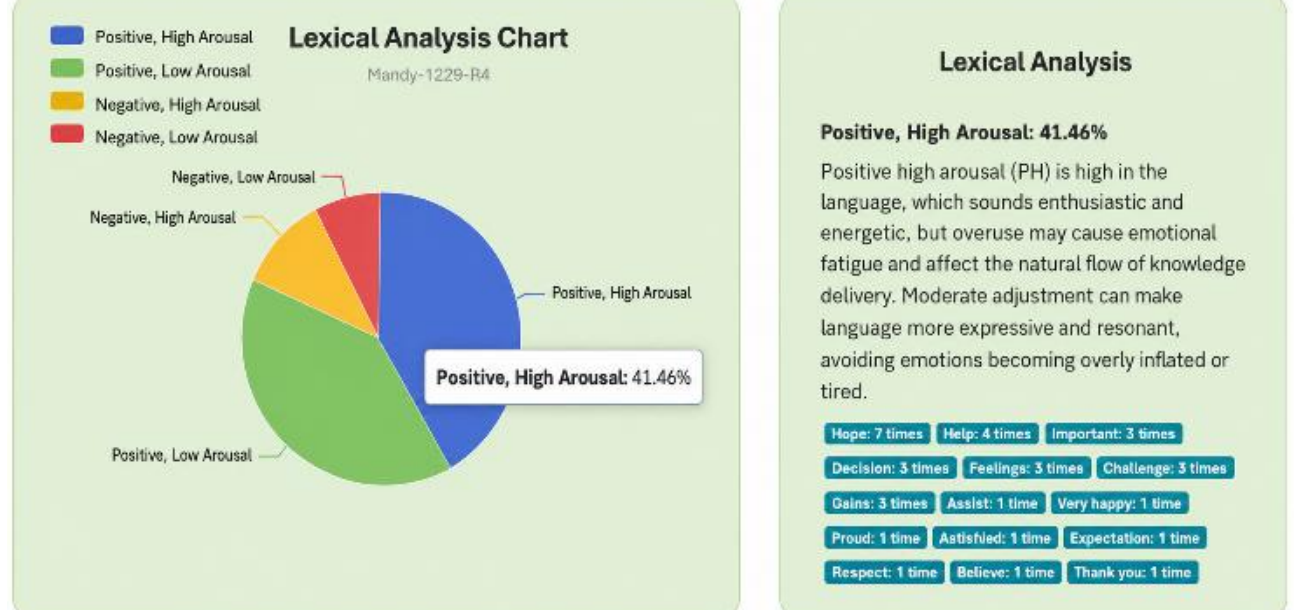

**Fig. 6. Lexical Emotive Expressions**

To translate interpretive insights into concrete revisions, the dashboard includes a rhetorical guidance module (Fig. 7). This module provides example-based script revisions generated using GPT-4.1 API under context engineering. For each instance, the system presents the learner's original wording, a revised version, and a brief explanation of the change. These targeted examples help learners refine tone, strengthen persuasive intent, and produce easier scripts to deliver on camera, thereby supporting improvement in Content and Clarity.

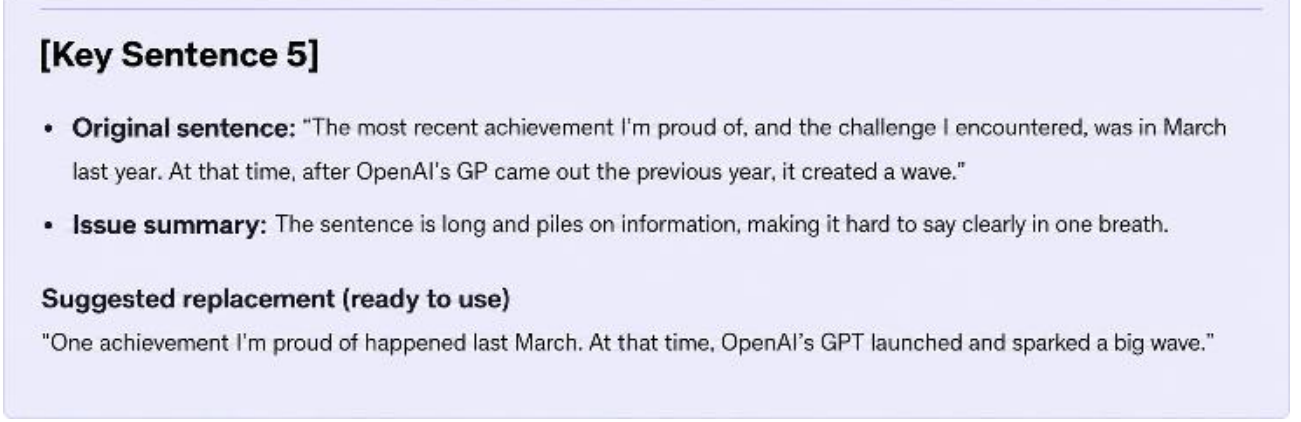

**Fig. 7. Rhetorical Guidance**

2) **Conversational AI Tutor**

The Conversational AI Tutor adds to the dashboard functionality by using interpreted analytics to inform practical, learner-centric advice. Whereas the dashboard helps the learner evaluate their presentation skills using BARS metrics and affective analyses, the tutor fills the loop by responding to learner inquiries and assisting in practice decision-making (**Fig. 8**).

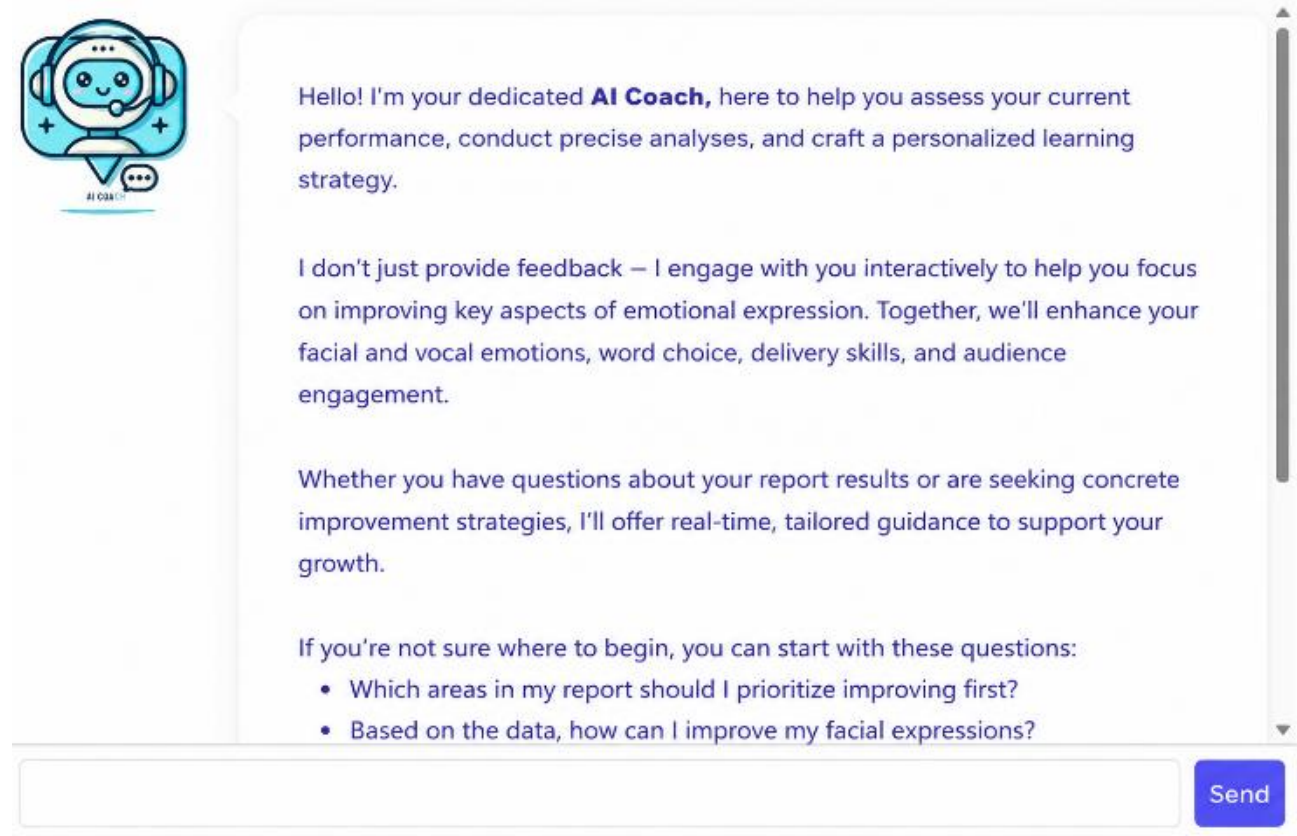

**Fig. 8. Conversational AI Tutoring**

The tutor is implemented as a retrieval-grounded conversational module built on GPT-4.1 with an RAG layer. Each response is anchored in analytic evidence retrieved from the learner's multimodal records, including transcripts, BARS-aligned scores, emotion descriptors, and multimodal feature summaries. The system architecture supports longitudinal reflection by maintaining learner records across practice submissions, which can be retrieved by the RAG layer to inform subsequent tutoring responses.

Beyond prompt configuration, we organized the tutor as a small library of modular tutoring routines, each maintained as a lightweight specification. Each routine defines a fixed instructional goal, a predefined set of required evidence fields from the learner record, and an output schema that constrains response structure, thereby standardizing pedagogy and reducing stylistic drift.

The RAG layer restricts generation to retrieved sources derived from the learner's own records, minimizing unconstrained free-form output. The tutor currently provides three modular routines: (1) performance summarization, (2) strength and weakness highlighting, and (3) translation of analytic indicators into learner-facing concepts to support subsequent self-directed practice. To keep feedback consistent with effective coaching, the phrasing and examples used by these routines draw on a curated library of expert coaching compiled from prior presentation training sessions with communication experts nominated by our MOU partners. All comments were screened to ensure alignment with the BARS rubric and the instructional logic of the GROW model, covering Goals for performance clarification, Reality for evidence-based assessment, Options for improvement identification, and Will for action-oriented commitments [45].

In such an integrated approach, users can form context-driven questions such as 'Why did I get a low score in Voice & Talk?' or 'How do I need to work on eye contact in a presentation?' In response to such inquiries, the teacher synthesizes learning insights from various levels of analysis, referring to the relevant views on the dashboard. Thus, fixed analytics are translated into an adaptive, interactive learning process by the dashboard.

A pilot study evaluated whether the tutor's feedback quality was comparable to professional communication coaching. The tutor generated 30 full dialogue sessions based on dashboard outputs. Each session was randomly assigned to three independent experts, who rated whether the feedback met coaching standards on a 5-point Likert scale (1 = strongly disagree, 5 = strongly agree). Agreement among raters was substantial (Krippendorff's $\alpha = .81$), which was computed on the 5-point ratings with an ordinal difference function. Session-level mean ratings suggested high perceived quality ($M = 4.10$, $SD = 0.38$).

### *E. Empirical Validation*

The system was examined in a 30-day field study centered on self-presentation practice. Participants were adult learners recruited from partner MOOC platforms who were enrolled in an oral presentation learning program. This sample was independent from the instructor cohort used for model development. Participants provided informed consent through the MOOC platform. All videos and log records were de-identified before analysis, and the research team did not access directly identifying information; data were stored and analyzed in secure environments under the platform's data governance procedures.

At baseline, participants submitted a pretest self-presentation video on a topic of their own choosing. The same topic was required for the posttest to ensure within-participant comparability. Following submission, participants gained access to the feedback dashboard and AI tutoring system. Over the subsequent 30 days, they reviewed analytic feedback and interacted with the AI tutor while refining their presentation delivery. Additional evaluative uploads were disabled during this period to encourage distributed practice. At the end of the window, participants submitted a single posttest self-presentation video under the same topical focus.

Inclusion in the analytic sample required both (a) verifiable exposure to the tutoring intervention between the pretest and posttest submissions and (b) valid automated scoring for both videos. Verifiable exposure was operationalized as at least one reliably logged conversational interaction with the AI tutor and a cumulative non-upload usage duration of at least five minutes, treated as a minimum threshold to ensure completion of at least one feedback interaction cycle rather than an indicator of learning intensity. Cases with no verifiable exposure were excluded. In addition, cases in which either the pretest or posttest video could not be processed for automated scoring due to file corruption or insufficient audiovisual quality were removed. Applying these criteria resulted in the exclusion of 36 cases, yielding a final analytic sample of $N = 204$.

Pretest videos averaged 202 seconds ($SD = 59.02$), and posttest videos averaged 210 seconds ($SD = 60.18$), with all recordings constrained to approximately 2–5 minutes. While presentation topics varied across participants, the task structure and topical consistency were held constant to reflect authentic self-presentation practice.

All videos were processed using an automated multimodal pipeline that generated BARS-aligned presentation scores and emotion-related expressive descriptors. Outputs were normalized at the segment level before aggregation to account for variation in recording length and speaking rate. Learning outcomes were evaluated by comparing pretest and posttest BARS scores, with effect sizes reported to quantify changes in presentation performance. Facial, vocal, and lexical emotion patterns were examined descriptively to characterize expressive adjustments associated with sustained system use.

## IV. RESULTS AND EVALUATION

### *A. Modeling Performance*

Before modeling, we examined the psychometric quality of the expert-annotated BARS ratings to ensure a reliable supervision signal. Inter-rater reliability was strong, with ICC(1,k) ranging from 0.78 to 0.89 across dimensions [46]. Overall Rating also aligned closely with the analytic dimensions (Pearson's $r = .85$; Spearman's $\rho = .83$), indicating coherent rubric structure and supporting the use of dimension-level scores as actionable reference points for feedback and iterative improvement.

Using these validated annotations, seven independent regression models were trained, one for each BARS dimension, using a multimodal feature representation with a total dimensionality of 4,374. Facial expression and head-movement information was encoded using a FaceMesh-based three-dimensional convolutional pipeline, producing 3,780 dynamic facial features computed over fixed 2-second temporal windows. Oculomotor behavior was represented by seven indicators capturing gaze stability, fixation dispersion, and directional variability.

Acoustic features consisted of 203 prosodic and spectral descriptors, including Mel-frequency cepstral coefficients (MFCCs), fundamental frequency (F0), and related summary statistics. Semantic content was modeled using sentence-level embeddings generated by a sentence-transformer model, yielding a 384-dimensional representation for each segment.

The dimensionality of each modality reflects the native output of its respective feature extraction pipeline rather than manual feature expansion. Because the resulting feature space is uneven across modalities, interpretability analyses were conducted at the modality level by aggregating feature attributions within each modality.

The dataset was partitioned by speaker into three disjoint subsets: training (70%), validation (15%), and test (15%). Hyperparameters were optimized on the training set using 5-fold speaker-independent cross-validation with Bayesian optimization [40]. Model selection was guided by performance on the validation set. The final models were retrained on the combined training and validation data and evaluated on the held-out test set.

As shown in **Table II**, the seven XGBoost models demonstrated moderate-to-strong agreement with expert ratings. Across the dimensions, $R^2$ values ranged from 0.48 to 0.61, Spearman's ρ fell between 0.69 and 0.78, and MAE values were generally between 0.43 and 0.57. These results indicate that the models capture rubric-aligned performance patterns with consistent rank-order agreement against expert annotations.

**Table II**. Multimodal XGBoost Model Performance

| Dimension | $R^2$ | MAE | ρ |
|---|---|---|---|
| Topic | 0.48 | 0.57 | 0.69 |
| Content | 0.50 | 0.55 | 0.71 |
| Clarity | 0.56 | 0.48 | 0.74 |
| Voice & Talk | 0.61 | 0.43 | 0.78 |
| Eye Contact | 0.53 | 0.52 | 0.72 |
| Nonverbal Expressiveness | 0.58 | 0.46 | 0.76 |
| **Overall Rating** | **0.59** | **0.44** | **0.77** |

Spearman's ρ of approximately 0.70 indicates meaningful ordinal agreement in behavioral scoring tasks, where expert judgments are inherently subjective. This level of correlation suggests alignment between model outputs and expert assessments.

To contextualize these results, we also trained unimodal baselines using only acoustic, facial, oculomotor, or textual features. Their performance was noticeably weaker. Across the seven dimensions, unimodal $R^2$ values mostly fell between 0.12 and 0.32, with Spearman correlations around 0.34 to 0.56. None of the single-modality models approached the consistency observed in the multimodal setting. This gap suggests that no individual signal source captures enough of the behavioral information that expert raters draw on, reinforcing the value of integrating multiple modalities for reliable scoring.

*B. Modeling Interpretability*

We examined how the models handled different types of signals by analyzing their Shapley Additive Explanations (SHAP) attributions. For each regressor, we computed the mean absolute SHAP value for every feature and aggregated these values by modality. To facilitate interpretation across dimensions, the modality-level SHAP values were normalized within each model, such that the values in **Table III** represent the relative contribution of each modality. Larger values indicate greater reliance on that signal type in the model's predictions.

**Table III**. SHAP-Based Modality Contributions

| Dimension | Acoustic | Facial | Oculomotor | Textual |
|---|---|---|---|---|
| Topic | 0.000 | 0.000 | 0.182 | 0.818 |
| Content | 0.000 | 0.000 | 0.173 | 0.827 |
| Clarity | 0.621 | 0.167 | 0.000 | 0.212 |
| Voice & Talk | 0.742 | 0.133 | 0.000 | 0.125 |
| Eye Contact | 0.000 | 0.000 | 0.801 | 0.199 |
| Nonverbal Expressiveness | 0.214 | 0.486 | 0.089 | 0.211 |
| **Overall Rating** | 0.338 | 0.352 | 0.110 | 0.200 |

Despite the high dimensionality of facial features (3,780-D), SHAP analysis revealed that modality importance is not proportional to feature count. For Voice & Talk, the 203 acoustic features collectively contributed 74.2% of predictive power, indicating that fewer, highly informative features can outweigh larger but more redundant feature sets.

When inspecting these values, several consistent patterns emerged. The models for Voice & Talk and Clarity showed higher relative reliance on acoustic information, which aligns with the evaluative emphasis of these dimensions in the BARS framework. The Eye Contact model relied predominantly on oculomotor features associated with gaze behavior. Textual features contributed most strongly to Topic and Content, while facial cues were the primary contributors for Nonverbal Expressiveness. Together, these patterns suggest that the models prioritize signal types consistent with the behavioral focus of each dimension, thereby supporting the interpretability of the feedback presented in the dashboard and the AI tutor.

*C. Empirical Validation and Evaluation*

This study employed a pre–post design with 204 adult learners to examine the pedagogical effects of the ITS using ML-generated BARS scores. Most participants held a bachelor's degree (61.76%), while the rest held a master's degree (38.24%). The sample included 38.24% women and 61.76% men.

Participants spanned a broad age range. Those aged 18–19 accounted for 2.94% of the sample, followed by 44.12% aged 20–25, 5.88% aged 26–30, 8.82% aged 31–35, 20.59% aged 36–40, 11.76% aged 41–45, and 5.88% aged 46 years and above. Disciplinary backgrounds comprised 64.71% participants from the social sciences and 35.29% from science-related fields.

At the time of participation, 70.59% were employed professionals and 29.41% were students in higher education. Reported work experience reflected this mix: 38.24% reported more than 10 years of experience, 8.82% reported 6–9 years, 2.94% reported 4–5 years, 32.35% reported 1–3 years, and 17.65% reported no formal work experience.

System logs, excluding video upload activity, showed an average usage count of 4.26 sessions (SD = 2.36, range = 1–8). Cumulative usage time averaged 35.79 minutes (SD = 13.01), with observed values ranging from 5.02 to 68.00 minutes.

As shown in **Table IV**, oral presentation performance was higher at posttest than at pretest across all assessed dimensions. Changes were observed in both structure-related aspects of presentation and delivery-related elements, with the latter showing more noticeable shifts. Improvements in voice use, eye contact, and nonverbal expressiveness contributed to the increase in Overall Rating.

**Table IV**. Paired Samples T-Test Results

| Dimension | t | Cohen's d | 95% CI for d Lower | Upper |
|---|---|---|---|---|
| Topic | 8.60 | 0.60 | 0.45 | 0.75 |
| Content | 5.56 | 0.39 | 0.25 | 0.53 |
| Clarity | 8.73 | 0.61 | 0.46 | 0.76 |
| Voice & Talk | 10.23 | 0.72 | 0.56 | 0.87 |
| Eye Contact | 7.40 | 0.52 | 0.37 | 0.66 |
| Nonverbal Expressiveness | 11.67 | 0.82 | 0.66 | 0.98 |
| **Overall Rating** | 12.83 | 0.90 | 0.74 | 1.06 |

To examine posttest performance while accounting for baseline differences, we estimated a multiple regression model with posttest Overall Rating as the outcome. Predictors included pretest Overall Rating, two practice indicators from system logs (usage frequency and cumulative non-upload time), and six demographic covariates, with categorical variables dummy coded using a reference group. The model was statistically significant, $F(9, 194) = 37.72$, $p < .001$, explaining 63.6% of the variance in posttest scores (adjusted $R^2 = .62$). Pretest performance remained a reliable predictor ($\beta = .16$, $p = .002$), and usage frequency showed a strong positive association ($\beta = .90$, $p < .001$), whereas cumulative usage time was smaller and did not reach conventional significance ($\beta = -.11$, $p = .066$). The regression results across models are summarized in **Table V**.

**Table V**. Regression Results (Key Predictors)

| Model | Pretest (β) | Usage count (β) | Usage Time (β) | Adjusted $R^2$ |
|---|---|---|---|---|
| Topic | .65*** | .86*** | -.25*** | 0.77 |
| Content | .88*** | .70*** | -.13** | 0.78 |
| Clarity | .80*** | .71*** | -0.05 | 0.61 |
| Voice & Talk | .54*** | .84*** | -.14** | 0.78 |
| Eye Contact | .71*** | .50*** | -0.08 | 0.79 |
| Nonverbal Expressiveness | .46*** | .75*** | -.17** | 0.64 |
| **Overall Rating** | .16** | .90*** | -0.11 | 0.62 |

Note. Entries are standardized coefficients (β). ** $p < .01$, *** $p < .001$

Overall, posttest outcomes were most strongly associated with documented practice frequency beyond baseline ability. This pattern suggests that improvement is associated with repeated, feedback-guided engagement with the system rather than passive exposure. Parallel regression analyses across the six additional BARS dimensions showed similar positive associations between practice frequency and posttest performance. In addition, pre–post comparisons indicated significant improvements across all seven dimensions. The pattern is not limited to a single composite outcome. Instead, the results reflect broader changes in presentation behavior across dimensions.

In addition to rubric-based gains, emotion-related analyses indicated systematic changes in expressive behavior following system use. Paired-samples tests revealed reliable pre–post differences across facial, vocal, and lexical signals, encompassing both negative and high-arousal expressions reflected in facial cues, vocal delivery, and word usage (e.g., facial anger and surprise; vocal anger and happiness; affectively charged lexical patterns; all $p < .05$). These changes were not confined to any single emotion, suggesting potential regulation of emotional expression during presentation rather than isolated affective shifts. Shifts in affective and expressive indicators further suggest that learners actively adjusted delivery-related behaviors during repeated practice, supporting the closed-loop process of evidence, interpretation, and targeted revision.

Accordingly, the results indicate significant improvements in on-camera oral presentation performance and expressive flexibility following the implementation of the closed-loop feedback design. The XGBoost-based framework provides stable and interpretable estimates of multimodal presentation behavior, enabling the system to generate consistent feedback signals that can be used for iterative behavioral adjustment. These findings are consistent with recent work on AI-supported presentation training, where rubric-based feedback and adaptive learning mechanisms have been used to support structured, iterative skill development [47], [48].

Building on these outcomes, the emotion-aware expressive layer supported learners in interpreting diagnostic results by making underlying expressive patterns more understandable, while the conversational AI tutor further facilitated reflection on these interpretations and guided learners toward concrete improvement strategies. As a result, these layers enabled learners to move from performance awareness to interpretation and ultimately to actionable behavioral change, illustrating a structured feedback-to-action learning mechanism within the ITS. The present design emphasizes feedback-supported reflection and iterative behavioral adjustment rather than real-time monitoring, as the goal is to support interpretable learning processes rather than continuous behavioral capture.

Several limitations should be noted. First, the pre–post design without a concurrent control group limits causal attribution; observed gains may partially reflect maturation, repeated measurement, or concurrent learning activities, or content familiarity effects introduced by requiring participants to address the same topic at pretest and posttest during the 30-day window. Future studies could further disentangle topic familiarity from skill development.

Second, outcomes were derived from automated, rubric-aligned scoring rather than direct expert scoring at posttest. While this supports scalability, it constrains the validity boundary of inferences to the extent that automated scores capture the construct of interest. Prior work has reported potential discrepancies between AI-based scoring and human evaluation, suggesting that results should be interpreted with attention to their alignment with human judgment criteria [49].

Third, the current implementation summarizes multimodal signals instead of modeling their temporal dynamics. This design supports interpretability for feedback but may overlook finer-grained variation in presentation behavior.

Fourth, the system does not provide real-time feedback during live performance. While real-time multimodal systems are increasingly feasible, they involve trade-offs in computational cost, latency, and system stability. The asynchronous, upload-based design ensures reliable multimodal feedback and supports a structured feedback and practice cycle for examining feedback-driven behavioral adjustment. It also enables stable and interpretable feedback representations that can be consistently applied across iterations, although it may limit the immediacy of feedback delivery.

Fifth, the system is built around a BARS rubric designed for on-camera presentation, where observable expressive behaviors can be systematically linked to feedback. Adapting the system to a different rubric or domain would require re-annotation of a dataset aligned with the target behavioral criteria and retraining of the scoring models, introducing a cold-start dependency at the scoring-model level. This dependency is shared by rubric-grounded assessment systems more broadly. The contribution of the present work lies in demonstrating a replicable architectural pipeline rather than a universally transferable scoring model. The annotation protocol, feature extraction procedure, and model training process are documented to support future adaptation efforts.

Finally, the analytic sample required verifiable practice and processable videos, which may introduce selection effects and limit generalizability to learners with comparable engagement and recording conditions. This reflects the intended use context of voluntary, feedback-driven practice rather than a controlled experimental sample. Within these constraints, the findings provide internally consistent, field-based evidence of behavioral improvement under stable evaluation conditions. This pattern is also consistent with prior controlled research in presentation training; for example, [20] using a randomized controlled design, reported that structured practice with feedback significantly improves presentation performance. Future work should incorporate controlled comparisons and external validation with independent human ratings to further strengthen generalizability and causal interpretation. In addition, the integration of deterministic multimodal scoring and generative LLM-based tutoring may provide a basis for future neurosymbolic affective tutoring systems.

## V. Conclusion

This study presented an integrated ITS for on-camera presentation training that combines rubric-aligned multimodal assessment, interpretable modeling, and conversational feedback. Technically, the system extends multimodal learning analytics with an interpretable XGBoost framework that captures complex behavioral cues while remaining anchored to a BARS-based rubric, supporting psychometric consistency and actionable score interpretation. Empirically, learners demonstrated consistent pre–post improvements across presentation dimensions, with the greatest improvements in delivery-related skills. The primary contribution of this study lies in demonstrating how multimodal analytic outputs can be systematically transformed into observable behavioral change within an ITS through an integrated feedback architecture. Conceptually, the results illustrate a structured learning mechanism in which performance analytics support self-monitoring, an emotion-aware expressive layer provides interpretable diagnostic cues, and a conversational tutor translates analytic evidence into concrete reflection and subsequent behavioral adjustment. These components together operationalize multimodal analytics into stable and interpretable feedback representations that can be consistently applied across practice iterations. On a larger scale, this work shows how ITS design can transform analytic outputs into sustained behavioral change through structured and reusable feedback mechanisms.

## Acknowledgment

The authors thank the partnering MOOC platforms and participants for their contributions. The authors also acknowledge Chih-Wen Lin for assisting with data collection and pilot testing.

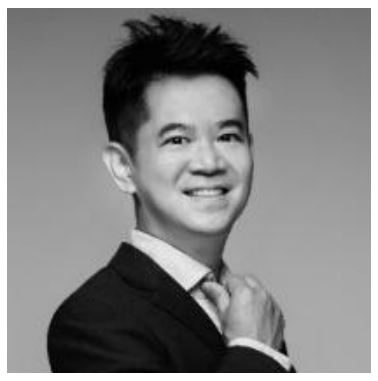

**Hung-Yue Suen** received the Ph.D. degree in Management Information Systems from National Chengchi University, Taipei, Taiwan. He is currently a Full Professor in the Department of Technology Application and Human Resource Development at National Taiwan Normal University, Taipei, Taiwan. His research interests include computational social systems, human–AI interaction, and learning technology.

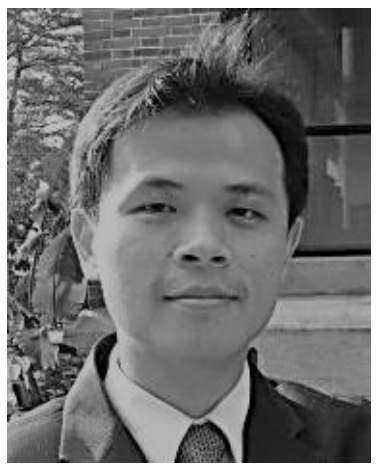

**Kuo-En Hung** received the Ph.D. degree in Human Resource Technology from National Taiwan Normal University, Taipei, Taiwan. He is currently an industrial researcher at National Taiwan Normal University. His research interests include computer vision, acoustic modeling, neural networks, and affective computing.